\documentstyle[aps,psfig,epsfig,multicol,eqsecnum]{revtex}

\begin{document}
\draft

\title{Quantum simulations of optical systems}
\author{M.Havukainen$^a$, G. Drobn\'y$^{b}$, S.Stenholm$^c$ and 
V.Bu\v{z}ek$^{b,d}$}
\address{
$^a$Helsinki Institute of Physics, P. O. Box 9, FIN-00014 University of 
Helsinki, Finland\\
$^b$Institute of Physics, Slovak Academy of Sciences, D\'{u}bravsk\'{a}
cesta 9, 842 28 Bratislava, Slovakia\\
$^c$Physics Department, Royal Institute of Technology, Stockholm, Sweden\\
$^d$Faculty of Mathematics and Physics,
Comenius University, Mlynsk\'{a} dolina F2, Bratislava, Slovakia
}

\date{\today}
\maketitle

\begin{abstract}
Within a framework of a two-dimensional
microscopic purely-quantum mechanical
model we analyze dynamics of single-photon wave packets
interacting with optical elements (beam splitters, mirrors) modeled
as systems of two-level atoms.
That is, we
utilize a two dimensional cavity to simulate the quantum behavior of
simple optical components and networks thereof. The field is quantized
using the canonical procedure, and only the basis states with one unit
of excitation are included. This, however, covers the linear optical
phenomena. The field is taken to interact with localized atoms through
a dipole interaction. Using different configurations of atoms and choosing
their frequencies to be resonant or off-resonance, we can model mirrors,
beam splitters, focusing devices and multicomponent systems. Thus we can
model arbitrary linear networks of optical components. We show the time
evolution of a photon wave packet
 in an interferometer as an example. As the state
of the field is known at each instant, spectral properties and spatial
coherence can immediately be obtained from the simulations. We also know
the states of the two level atoms constituting the components, which
allows us to consider their quantum behavior. Here the decay of an excited
atom into the vacuum state of the electromagnetic field in the
two-dimensional cavity is studied.
\end{abstract}




\begin{multicols}{2}

\section{Introduction}

It is well understood that the electromagnetic fields giving rise to all
optical phenomena have ultimately to be represented by quantum operators.
These couple to the degrees of freedom of matter and their modification due
to this interaction constitutes the quantum counterpart of the action of
optical components. Ordinary optical devices operate in the linear regime of
interaction, but the important area of Nonlinear Optics is based on higher
order effects of the field-matter coupling. In most situations, the optical
phenomena can be described entirely in terms of classical fields, but many
recent investigations require that the quantum character of the field is
accounted for. Such research constitutes the topics of Quantum Optics
\cite{Loudon,MilburnandWalls,Mandelfree}. 
However, many quantum effects are of interest even in
the linear regime of operation: quantum noise \cite{Gardiner}
sets the limit to
communication by optical channels and amplifiers, quantum interference shows
up in precision measurements and tests of fundamental issues and also in
reading and writing quantum information. Manipulation of quantum information
such as quantum computations usually requires the inclusion of higher order
effects, i.e. nonlinear interactions between the qubits \cite{Steane}.

In this paper, we are going to discuss the dynamics of single photon 
wave packets 
in various two-dimensional atomic configurations. These are taken to be
models of optical networks, where we explicitly include the atomic nature of
the optical components distributed over the volume under investigation. 
This approach provide us with a completely microscopic quantum-mechanical
picture of how photon wave packets interact with optical elements
represented as collections of two level atoms.
For
practical reasons, we have to restrict our work to one-photon states, but
this is not such a serious limitation as it may seem. All {\em linear} optical
effects are based on the single-photon interacting with material structures,
and consequently we have a general description of Quantum Optics phenomena
in the linear regime. The need to consider multi-photon effects arises only
in connection with the quantum treatment of Nonlinear Optics.

There are two basic ways to approach the quantization of optical systems. In
the conventional one, we determine a complete set of eigenmodes of the total
universe, and express the fields of interest in terms of these. Any matter
present is described through its interaction with the fields, and the
coupled field-matter problem is then solved to the best of our ability. This
is the approach utilized in traditional QED, and its development is found in
many standard texts. The alternative approach, designed for Quantum Optics
applications, is to determine the eigenmodes of the system at the classical
level, the matter involved is then treated as boundary conditions on the
field modes. Especially the new area of Cavity QED research \cite{berman},
utilizes this point of view, and it provides the basis both for quantum
communication theories and many fundamental investigations. 

In the field of optics, the components are usually treated as boundary
conditions only, and the complete optical device is considered to be an
optical network. This approach has been discussed thoroughly in the
classical regime of operation \cite{weissman}. For linear devices, the
classical treatment can be taken over into the quantum regime by the use of
suitable Quantum Optics tools \cite{ekert,stenholm94,torma}.
In principle, any device understood classically, can
be treated quantum mechanically with such an approach. The specific quantum
features manifest themselves in the initial conditions and the restrictions
on observability imposed by quantum theory \cite{stenholm95}.

Another specifically quantum mechanical effect is the occurrence of
spontaneous decay. Within a one-dimensional model of the modes of the
universe, this is discussed in Ref. \cite{buzek2}, where both free
Weisskopf-Wigner decay and cavity modified decay are discussed. Such
phenomena have been the object of much interest within QED research, for an
extensive list of references see Ref. \cite{buzek2}. Within the model chosen
there, one can see the emergence of the exponential law and the inhibition
of decay observed in a photonic band gap structure. In general, the model
provides insight into the role of atomic media in the irreversible transfer
of excitation energy into the field modes of the universe.

In this paper we combine the two views discussed above: We retain a
description in terms of a complete set of two-dimensional eigenmodes of the
universe. The optical components are described in terms of their atomic
constituents. All atomic structures are represented by spatially localized
two-level atoms. These are treated as point-like partciles 
in accordance with the
dipole approximation assumed to be valid. The state of the field is taken to
be a single photon wave packet with a narrow energy distribution. In this case,
the state can be described by a truncated expansion in terms of modes of the
universe. The spatially distributed two-level atoms describing the
structures are taken to be initially in their ground states. The atoms 
can be chosen
to resonate with the central frequency of the 
photon wave packet 
or be well off resonance; various effects can be
modeled in this way. When the single photon is absorbed, only one of the
atoms is excited, and the field is reduced to its ground state. Such a
choice limits the Hilbert space needed in the calculations to manageable
size, but allows us to investigate many simple networks of significance in
linear optics. All such effects are, in principle, describable at the single
photon level; only Nonlinear Optics effects require more photons, which
would make the Hilbert space expand beyond the limits of available computer
resources.

Our approach based on a complete set of eigenmodes allows us to investigate
the dynamic performance of many linear systems. In order to illustrate the
method, we select the simplest optical components: mirrors, beam splitters,
focusing devices and interferometers. The over-all performance of the
components follows directly from their classical theory, but our approach
allows us to investigate the microscopic (quantum) 
behavior of the setup. Quantum
coherence between various spatial regions in the device is directly visible
in the states calculated, and the time and space scales of the various
interferometric structures can be read off the results. Combined with
various models of measurements, our calculations contain considerably more
information than a simple classical computation. Here we only discuss the
measurement of frequency and the possible occurrence of filtering action in
the atomic structures, which does not in itself depend too much on the
quantum nature of the fields. But modeling the frequency detection by
atomic absorption, we utilize the full character of the model, which allows
further extension to quantum correlation measurements if we so desire.

Our work is based on the model put forward in \cite{buzek1} which we
extend to two dimensions. The quantized modes of the universe are introduced
in Sec. II together with their interaction with the spatially distributed
atoms. In Sec. III we specify the details of the model and indicate how the
calculations have been carried out. Section IV presents the various simple
components analyzed in this paper. We describe how they are modeled and
show the results of the detailed solution of the time evolution. Finally in
Sec. V we present our conclusions and discuss possible extensions and
applications of the work.

\section{Operators for the free field in two dimensions}

The field is enclosed inside a two dimensional cavity determined by the 
relations

\begin{equation}
-\frac{L}{2}\leq x,y \leq \frac{L}{2}.
\end{equation}
The periodic boundary conditions restrict the allowed values in 
${\bf k}$-space to a
 discrete set

\begin{equation}
\label{kvalues}
k_i=\frac{2\pi n_i}{L},  \ \ \ i=x,y.
\end{equation}
In computer simulations, the ${\bf k}$-values must be restricted by giving some 
upper limit for the integer $n$ which corresponds to a specific frequency
cut-off.
The electric and magnetic field can be expanded \cite{Mandelfree} using the mode 
functions

\begin{eqnarray}
\label{efree}
& & \hat{\bf E}({\bf r}) = \frac{i}{L}\sum_{{\bf k}s}
\left(\frac{\hbar\omega_{{\bf k}s}}{2\epsilon_0}\right)^{1/2}
(\hat{a}_{{\bf k}s}{\bf \epsilon}_{{\bf k}s}e^{i{\bf k}\cdot{\bf r}} - 
{\rm h.c}) \\
& & \hat{\bf B}({\bf r}) = \frac{i}{L}\sum_{{\bf k}s}
\left(\frac{\hbar}{2\epsilon_0\omega_{{\bf k}s}}\right)^{1/2}
(\hat{a}_{{\bf k}s}({\bf k}\times{\bf \epsilon}_{{\bf k}s})e^{i{\bf k}
\cdot{\bf r}} 
 - {\rm h.c}),
\label{bfree}
\end{eqnarray}
where the summation $\sum\limits_{{\bf k}s}$ is over all ${\bf k}$-values 
(\ref{kvalues}) and two polarization
indices $s=1,2$. The frequency $\omega_{{\bf k}s}$ is the same for both 
polarizations

\begin{equation}
\label{frequency}
\omega_{{\bf k}s} = c|{\bf k}|.
\end{equation}
The general {\bf k}-vector in two dimensions can be written

\begin{equation}
{\bf k} = k_x\hat{e}_1 + k_y\hat{e}_2 = |{\bf k}|(\cos(\phi)\hat{e}_1 + 
\sin(\phi)\hat{e}_2).
\end{equation}
The polarization vectors which obey the usual right hand rule conventions are
\begin{eqnarray}
& & {\bf \epsilon}_{{\bf k}1} = -\hat{e}_3 \\
& & {\bf \epsilon}_{{\bf k}2} = -\sin(\phi)\hat{e}_1 + \cos(\phi)\hat{e}_2.
\end{eqnarray}
The ${\bf k}$-vector and polarization indexes satisfy the relations 
\cite{Mandelfree}

\begin{eqnarray}
& & {\bf\epsilon}_{{\bf k}i}\cdot{\bf\epsilon}_{{\bf k}j}=\delta_{ij} \\
& & \sum\limits_{ss'}{\bf\epsilon}_{{\bf k}s}\cdot{\bf\epsilon}_{{\bf k}s'}=2\\
& & {\bf k}\times{\bf \epsilon}_{{\bf k}1} = -k_y\hat{e}_1 + k_x\hat{e}_2\\
& & {\bf k}\times{\bf \epsilon}_{{\bf k}2} = |{\bf k}|\hat{e}_3.
\end{eqnarray}
The energy-density operator is

\begin{equation}
\label{energydensity}
\hat{H}({\bf r}) = \frac{1}{2}\epsilon_0\hat{{\bf E}}^2({\bf r}) 
+ \frac{1}{2\mu_0}\hat{{\bf B}}^2({\bf r})
\end{equation}
Using (\ref{efree}) and (\ref{bfree}) gives

\end{multicols}

\widetext

\begin{eqnarray}
\frac{1}{2}\epsilon_0{\bf \hat{E}}^2({\bf r}) & = & -\frac{\hbar}{4L^2}
\sum\limits_{{\bf k}{\bf k}'ss'}\sqrt{\omega_{\bf k}\omega_{{\bf k}'}}\big(
\hat{a}_{{\bf k}s}\hat{a}_{{\bf k}'s'}e^{i{\bf k}\cdot{\bf r} + i{\bf k}'
\cdot{\bf r}}- \hat{a}_{{\bf k}s}\hat{a}_{{\bf k}'s'}^{\dag}e^{i{\bf k}
\cdot{\bf r} - i{\bf k}'\cdot{\bf r}} \\ \nonumber
& & \hspace{5cm}-\hat{a}_{{\bf k}s}^{\dag}\hat{a}_{{\bf k}'s'}e^{-i{\bf k}\cdot{\bf r} 
+ i{\bf k}'\cdot{\bf r}}
+\hat{a}_{{\bf k}s}^{\dag}\hat{a}_{{\bf k}'s'}^{\dag}e^{-i{\bf k}\cdot{\bf r} 
- i{\bf k}'\cdot{\bf r}}\big) \\
\frac{1}{2\mu_0}{\bf \hat{B}}^2({\bf r}) 
& = & -\frac{\hbar}{4L^2\mu_0\epsilon_0}
\sum\limits_{{\bf k}{\bf k}'ss'}\frac{1}{\sqrt{\omega_{\bf k}\omega_{{\bf k}'}}}
(\hat{a}_{{\bf k}s}\hat{a}_{{\bf k}'s'}e^{i{\bf k}
\cdot{\bf r} + i{\bf k}'\cdot{\bf r}}
- \hat{a}_{{\bf k}s}\hat{a}_{{\bf k}'s'}^{\dag}e^{i{\bf k}\cdot{\bf r} - 
i{\bf k}'\cdot{\bf r}}\\ \nonumber
& & -\hat{a}_{{\bf k}s}^{\dag}\hat{a}_{{\bf k}'s'}e^{-i{\bf k}\cdot{\bf r} 
+ i{\bf k}'\cdot{\bf r}}
+\hat{a}_{{\bf k}s}^{\dag}\hat{a}_{{\bf k}'s'}^{\dag}e^{-i{\bf k}\cdot{\bf r} 
- i{\bf k}'\cdot{\bf r}})
[(k_xk_x' + k_yk_y')\delta_{s1}\delta_{s'1} + |{\bf k}||{\bf k}'|\delta_{s2}\delta_{s'2}]
\end{eqnarray}

\begin{multicols}{2}

In our simulations, we have restricted the polarization of the field to 
${\bf\epsilon}_{{\bf k}1}$.
The modes with $s=2$ are taken to have have zero amplitudes. In addition to that we 
restrict the number of excitations
of our basis vectors to one. For these kind of basis vectors the terms
$\hat{a}_{{\bf k}s}\hat{a}_{{\bf k}'s'}$ and 
$\hat{a}_{{\bf k}s}^{\dag}\hat{a}_{{\bf k}'s'}^{\dag}$
do not give any contribution. These terms can be omitted from the expressions. 
For the states described
above, the expectation
values are obtained by replacing the operators with the coefficients of the 
corresponding statevectors $\hat{a}_{\bf k} \rightarrow c_{\bf k}$ and
$\hat{a}_{\bf k}^{\dag} \rightarrow c_{\bf k}^*$. The normally-ordered terms 
in the energy density become (normal ordering is indicated by colons)

\begin{eqnarray}
:\frac{1}{2}\epsilon_0{\bf \hat{E}}^2({\bf r}): & = & \frac{\hbar}{2L^2}RR^* \\
:\frac{1}{2\mu_0}{\bf \hat{B}}^2({\bf r}): & = & 
\frac{\hbar}{2L^2\epsilon_0\mu_0}(S_xS_x^* + S_yS_y^*),
\end{eqnarray}
where

\begin{eqnarray}
R & = & \sum\limits_{\bf k}\sqrt{\omega_{\bf k}}c_{\bf k}
e^{i{\bf k}\cdot{\bf r}} \\
S_i & = & \sum\limits_{\bf k}\frac{k_i}{\sqrt{\omega_{\bf k}}}c_{\bf k}
e^{i{\bf k}\cdot{\bf r}},\ \ \ i=x,y.
\end{eqnarray}
The two-fold summation over the ${\bf k}$-space is seen to factorize and the 
formulas for $R$ and
$S_i$ are Fourier transforms of two different functions. 
For numerical simulations these two properties
are essential as will be seen later.
We note that if the polarization is such that the modes with $s=1$ are taken
to have zero amplitudes, then the two terms 
$:\frac{1}{2}\epsilon_0{\bf \hat{E}}^2({\bf r}):$
and 
$:\frac{1}{2\mu_0}{\bf \hat{B}}^2({\bf r}):$ 
in the expression for the energy density are equal.

Integrating (\ref{energydensity}) over the spatial coordinates and using the 
integral

\begin{equation}
\int\limits_{-L/2}^{L/2}dx\int\limits_{L/2}^{L/2}dye^{i({\bf k} - {\bf k}')
\cdot{\bf r}}=L^2\delta_{{\bf k}{\bf k}'}
\end{equation}
gives the familiar form

\begin{equation}
\label{hf}
\hat{H}_F = \frac{1}{2}
\sum\limits_{{\bf k}}\hbar\omega_{{\bf k}}
(\hat{a}_{{\bf k}}^{\dag}\hat{a}_{{\bf k}} +
\hat{a}_{{\bf k}}\hat{a}_{{\bf k}}^{\dag})=
\sum_{{\bf k}} \hbar\omega_{\bf k}(\hat{a}_{{\bf k}}^{\dag}\hat{a}_{{\bf
k}}+\frac{1}{2}),
\end{equation}
which in the normally ordered form reads 
$:\hat{H}_F: = 
\sum\limits_{{\bf k}} \hbar\omega_{\bf k}\hat{a}_{{\bf k}}^{\dag}\hat{a}_{{\bf k}}$

\section{The General Hamiltonian and the states}

In the previous chapter the formulas for the field in the vacuum were derived. 
In this chapter we add
an assembly of two level atoms to the cavity and give the corresponding 
Hamiltonians. The general form of the
statevector with one excitation is also given. The material presented here 
is based on the
similar simulations in one dimension done by V.Bu\v{z}ek et.al. 
\cite{buzek2}.
The simulations in two dimensions are numerically more demanding, 
but we have been able to develop
efficient numerical methods which make these simulations possible.

\subsection{The Hamiltonian}

The total Hamiltonian $\hat{H}$ can be divided into three parts

\begin{equation}
\hat{H}=\hat{H}_F + \hat{H}_A + \hat{H}_I,
\end{equation}
where the field Hamiltonian is given by equation (\ref{hf}). The atomic 
Hamiltonian is  the sum over all one-atom Hamiltonians

\begin{equation}
\hat{H}_A=\sum\limits_{j=1}^{N_A}\hbar\omega_j\hat{\sigma}_z^j
\end{equation}
where $\omega_j$ is the transition frequency 
of the $j$-th atom and $\hat{\sigma}_z^j$ is Pauli's spin matrix.
In the interaction Hamiltonian the dipole approximation is used. For simplicity the dipole
operator is taken to be

\begin{equation}
{\bf \hat{D}}_j = (D_j\hat{\sigma}_+^j + D_j^*\hat{\sigma}_-^j)\hat{e}_3,
\end{equation}
i.e. it has a component in the $\hat{e}_3$ direction only. The general dipole
vector would have components
in the $x$- and $y$-directions too. The interaction Hamiltonian has the form

\begin{equation}
H_I =-\sum\limits_{j=1}^{N_A}\hat{\bf D}_j\cdot\hat{\bf E}({\bf r}_j),
\end{equation}
where $\hat{\bf E}({\bf r}_j)$ is the electric field operator (\ref{efree}) at
the position of the atom.
The rotating wave approximation (RWA) is to be used, and we neglect the
$\hat{\sigma}+^j\hat{a}^{\dag}$ - and
$\hat{\sigma}_-^j\hat{a}$ terms. 
In addition to that we replace the mode frequency in the electric field 
operator by the atomic
frequency and use the dot products 
$\hat{e}_3\cdot{\bf\epsilon}_{{\bf k}1}=-1$ and 
$\hat{e}_3\cdot{\bf\epsilon}_{{\bf k}2}=0$ to get

\begin{equation}
\label{hi}
\hat{H}_I \equiv \hat{H}_{I1} + \hat{H}_{I2} = \sum\limits_{j=1}^{N_A}
\sum\limits_{{\bf k}} \left( g(j,{\bf k})\hat{\sigma}_+^j\hat{a}_{{\bf k}}g^*(j,{\bf k})\hat{\sigma}_-^j\hat{a}_{{\bf k}}^{\dag} \right),
\end{equation}
 in what follows we omit the polarization index in subscripts of field
operators. The coupling constant is

\begin{equation}
\label{couplingconstant}
g(j,{\bf k})=-\frac{i\hbar}{2\epsilon_0L}\sqrt{\omega_{\bf k}}D_j 
e^{i{\bf k}\cdot{\bf r}_j}.
\end{equation}
Only those modes whose resonance frequency is close to the atomic frequency 
interact significantly
with the atom, so we can replace the mode frequency $\omega_{\bf k}$ by the 
atomic frequency
$\omega_j$ in equation (\ref{couplingconstant}).

\subsection{The statevector}

In all simulations we have restricted the total number of excitations to one. 
Consequently,  
the most general statevector of the atom-field system   has the form

\begin{eqnarray}
\label{statevec}
|\Psi\rangle & = & \sum\limits_{\bf k}\left( c_{\bf k}|1\rangle_{\bf k}
\prod\limits_{{{\bf k}'}\neq {\bf k}}\right) 
|0\rangle_{{\bf k}'}\otimes\prod\limits_{j=1}^{N_A}|0\rangle_j \nonumber \\
& & + \sum\limits_{\bf k}|0\rangle_{\bf k}\otimes
\sum\limits_{j=1}^{N_A}\left( c_j|1\rangle_j
\prod\limits_{j'=1,j'\neq j}^{N_A}|0\rangle_{j'}\right) \\ 
& \equiv & \sum\limits_{\bf k}c_{\bf k}|1_{\bf k},\{0\}\rangle 
+ \sum\limits_{j=1}^{N_A}c_j|\{0\},1_j\rangle.    \nonumber
\end{eqnarray}
The first sum contains all the basis vectors where the excitation is in 
one of the field modes and all
the atoms are in the ground state. In the second sum the field modes 
are in the vacuum state and one of
the atoms is excited. The complex numbers $c_{\bf k}$ and $c_j$ are 
the probability amplitudes of the
corresponding basis vectors. We have dropped the polarization indices 
because in our simulations 
only the basis vectors
with the polarization vector ${\bf\epsilon}_{{\bf k}1}$ are excited as was 
discussed earlier.

The general Gaussian one photon statevector is of the form

\begin{equation}
|\Psi\rangle =\sum\limits_{\bf k}c_{\bf k}|1_{\bf k},\{0\}\rangle,
\end{equation}
where the mode coefficient $c_{\bf k}$ is

\begin{eqnarray}
\label{generalGauss}
\lefteqn{c_{\bf k}=\frac{e^{i{\bf k}\cdot{\bf r}_0}}{\sqrt{4\pi^2M}}\exp\big(-\frac{\Delta_{ky}^2}{4M}(k_x-k_{x0})^2}  \\
& & - \frac{
\Delta_{kx}^2}{4M}(k_y-k_{y0})^2 + \frac{\Delta_{kx,ky}^2}{2M}(k_x-k_{x0})(k_y-k_{y0})\big). \nonumber
\end{eqnarray}
The parameters $M$ and $\Delta_{kx,ky}^2$ are

\begin{eqnarray}
M & = & \Delta_{kx}^2\Delta_{ky}^2 - (\Delta_{kx,ky}^2)^2 \\
\Delta_{kx,ky}^2 & = & \langle k_x k_y\rangle - \langle k_x\rangle\langle k_y\rangle.
\end{eqnarray}
If the cross-variance $\Delta_{kx,ky}^2$ vanishes the formula for $c_{\bf k}$ reduces
to two independent Gaussian distributions

\begin{eqnarray}
\label{Gauss}
\lefteqn{c_{\bf k}=(2\pi\Delta_{kx}^2)^{-1/4}
(2\pi\Delta_{ky}^2)^{-1/4}e^{-i{\bf k}\cdot{\bf r}_0}} \nonumber \\
& & \exp\left[-\frac{(k_x - k_{x0})^2}{4\Delta_{kx}^2}-
\frac{(k_y - k_{y0})^2}{4\Delta_{ky}^2}\right]
\end{eqnarray}
All initial distributions used in our simulations are of the form
(\ref{Gauss}).
The distribution (\ref{Gauss}) in ${\bf k}$-space is centered around $(k_{x0}, k_{y0})$
with the corresponding central frequency $\omega_0$.
If
$\Delta_{kx}^2=\Delta_{ky}^2$ the distribution is symmetric. If
$\Delta_{kx}^2<\Delta_{ky}^2$ the distribution is wider in the $y$-direction
(and vice versa). 
The variances in ${\bf k}$-space and configuration space are inversely
proportional. If $\Delta_{kx}^2$ is small, the energy density distribution in
configuration space is wide in the x-direction.
The normally ordered energy
distribution associated with the state (\ref{generalGauss}) or (\ref{Gauss})
is well localized near the point ${\bf r}_0$ in the configuration space. 
Essential for this is the phase part $e^{-i{\bf k}\cdot{\bf r}_0}$
of the coefficient $c_{\bf k}$. If the form of the phase was different
the intensity profile would not be Gaussian.

The time evolution of the Gaussian wave packet inside an empty cavity is 
determined by the Hamiltonian
$\hat{H}_F$ (\ref{hf}) with the corresponding 
evolution operator $\exp(-\frac{i}{\hbar}\hat{H}_Ft)$.
Applying
this to the state (\ref{statevec}) gives for the time-evolution of the 
coefficients
$c_{\bf k}(t) = c_{\bf k}(0)e^{-i\omega_{\bf k}t}$. The absolute value 
of the coefficients remain the
same, only the phase changes. For the phase part we get

\begin{equation}
\exp(-i{\bf k}\cdot{\bf r}_0 - i\omega_{\bf k}t) 
= \exp[-i{\bf k}\cdot({\bf r}_0 + ct{\bf e}_{\bf k})],
\end{equation}
where ${\bf k}=|{\bf k}|{\bf e}_{\bf k}$. 
The time evolution inside the empty cavity reduces to the time
evolution of the parameter ${\bf r}(t) = {\bf r}_0 + ct{\bf e}_{\bf k}$. 
We remember that the phase factor determine the shape of the normal
ordered intensity profile. Because the time evolution of the phase
is different for different modes, the normal ordered intensity does
not preserve its original Gaussian shape.
If the direction of the vector  ${\bf e}_{\bf k}$ is more or less
the same for all basis vectors which have nonzero coefficients, 
the shape of the energy density distribution remains 
approximately the same longer.
The situation is like this when the
state vector in ${\bf k}$-space is centered around some ${\bf k}$-value
far from the origin and  the variances are small.

\subsection{Transformation to the interaction picture}

It turned out to be faster to carry out the numerical integration in the
interaction picture.
The transformation Hamiltonian is $\hat{H}_0=\hat{H}_A + \hat{H}_F$. 
The interaction
Hamiltonian in the rotating frame is

\begin{equation}
\hat{H}_I^{(I)}=\exp(i\hat{H}_0t/\hbar)\hat{H}_I\exp(-i\hat{H}_0t/\hbar),
\end{equation}
which is obtained by the following replacement

\begin{eqnarray}
\hat{a}_{\bf k} & \rightarrow & \hat{a}_{\bf k}e^{-i\omega_{\bf k} t} \nonumber \\
\hat{a}_{\bf k}^{\dag} & \rightarrow & \hat{a}_{\bf k}^{\dag}e^{i\omega_{\bf k} t} 
\nonumber \\
\hat{\sigma}_{-}^j & \rightarrow & \hat{\sigma}_{-}^j e^{-i\omega_j t} \\
\hat{\sigma}_{+}^j & \rightarrow & \hat{\sigma}_{+}^j e^{i\omega_j t} \nonumber
\end{eqnarray}
in equation (\ref{hi}), and we get

\begin{eqnarray}
\label{hiinint}
\hat{H}_I^{(I)} & = & \hat{H}_{I1}^{(I)} + \hat{H}_{I2}^{(I)} \nonumber \\
& = & \sum\limits_{j=1}^{N_A}\sum_{\bf k}\big( g(j,{\bf k})
e^{i(\omega_j - \omega_{\bf k})t}\hat{\sigma}_+^j\hat{a}_{\bf k} \\
& & \hspace{1cm} + g^*(j,{\bf k})e^{-i(\omega_j - \omega_{\bf k})t}\hat{\sigma}_-^j
\hat{a}_{\bf k}^{\dag} \big). \nonumber
\end{eqnarray}
The statevectors in the interaction picture become

\begin{eqnarray}
\label{psitointact}
|\Psi\rangle^{(I)} & = & \exp(i\hat{H}_0t/\hbar)|\Psi\rangle \\
& = & \sum\limits_{\bf k}c_{\bf k}e^{i\omega_{\bf k}t}|1_{\bf k},\{ 0\} \rangle 
+ \sum\limits_{j=1}^{N_A}c_j e^{i\omega_j t}|\{ 0\} ,1_j\rangle . \nonumber
\end{eqnarray}
and the Schr\"odinger equation for the wavefunction is

\begin{equation}
\label{schrinint}
i\hbar\frac{d|\Psi\rangle^{(I)}}{dt} = \hat{H}_I^{(I)}|\Psi\rangle^{(I)}.
\end{equation}
Integration of the Schr\"odinger equation in the interaction picture 
is faster than
the original equation because only the interaction Hamiltonian is present.

\subsection{Numerical methods}

\subsubsection{Integration of Schr\"odinger equation}

Our choice for the integration method of the time dependent Schr\"odinger
equation is a classical four stage
fourth order Runge-Kutta method. If the wavefunction at time $t$ is
$|\Psi (t)\rangle$ the wavefunction
at a later time $t+\Delta t$ ($\Delta t$ small) $|\Psi (t)\rangle$ is given by
the following algorithm \cite{nrinc}

\begin{eqnarray}
\label{rkequations}
|k_1\rangle & = & \Delta t\hat{H}|\Psi(t)\rangle \nonumber \\
|k_2\rangle & = & \Delta t\hat{H}(|\Psi(t)\rangle + 0.5|k_1\rangle )\nonumber \\
|k_3\rangle & = & \Delta t\hat{H}(|\Psi(t)\rangle + 0.5|k_2\rangle ) \\
|k_4\rangle & = & \Delta t\hat{H}(|\Psi(t)\rangle + 0.5|k_3\rangle )\nonumber \\
|\Psi(t+\Delta t)\rangle & = &  |\Psi(t)\rangle + 
\frac{|k_1\rangle}{6} + \frac{|k_2\rangle}{3} + 
\frac{|k_3\rangle}{3} + \frac{|k_4\rangle}{6} + O((\Delta t)^5). \nonumber
\end{eqnarray}
The timestep $\Delta t$ is a fixed constant.

The essential part of the integration from the numerical point of view is 
how to evaluate
the right hand part of the equation (\ref{schrinint}) as efficiently 
as possible. The first
term in the equation (\ref{hiinint}) gives

\begin{eqnarray}
\label{hi11}
\lefteqn{\hat{H}_{I1}|1_{\bf k},\{0\}\rangle  = } \nonumber \\
& & -\frac{i\hbar}{2\epsilon_0L}\sum\limits_{j=1}^{N_A}
\sum\limits_{\bf k'}\sqrt{\omega_j}D_j\exp(i{\bf k'}
\cdot{\bf r}_j)e^{i(\omega_j-\omega_{\bf k})t}\sigma_+^j
\hat{a}_{{\bf k'}1}|1_{\bf k},\{0\}\rangle \nonumber \\
& & = -\frac{i\hbar}{2\epsilon_0L}\sum\limits_{j=1}^{N_A}
\sqrt{\omega_j}D_j\exp(i{\bf k}\cdot{\bf r}_j)
e^{i(\omega_j-\omega_{\bf k})t}|\{0\},1_j\rangle  \\
& & \hat{H}_{I1}|\{0\},1_j\rangle = 0,
\label{hi12}
\end{eqnarray}
and the second one

\begin{eqnarray}
\label{hi21}
& & \hat{H}_{I2}|1_{\bf k},\{0\}\rangle = 0 \\
\label{hi22}
\lefteqn{\hat{H}_{I2}|\{0\},1_j\rangle =} \\
& &  \frac{i\hbar}{2\epsilon_0L}
\sum\limits_{j=1}^{N_A}\sum\limits_{\bf k}
\sqrt{\omega_j}D_j^*\exp(-i{\bf k}\cdot{\bf r}_j)
e^{-i(\omega_j-\omega_{\bf k})t}\sigma_-^j
\hat{a}_{{\bf k}1}^{\dag}|\{0\},1_j\rangle \nonumber \\
& & =\frac{i\hbar}{2\epsilon_0L}\sum\limits_{\bf k}
\sqrt{\omega_j}D_j^*\exp(-i{\bf k}\cdot{\bf r}_j)
e^{-i(\omega_j-\omega_{\bf k})t}|1_{\bf k},0\rangle . \nonumber
\end{eqnarray}
Hence the new coefficients for the atomic ($c_j'$) and field ($c_{\bf k}'$) 
basis vectors become

\begin{eqnarray}
c_j' & = & -\frac{i\hbar}{2\epsilon_0L}\sqrt{\omega_j}
D_j e^{i\omega_j t}T({\bf r}_j,t) \\
c_{\bf k}' & = & \frac{i\hbar}{2\epsilon_0L}e^{i\omega_{\bf k}t}U({\bf k},t),
\end{eqnarray}
where

\begin{eqnarray}
& & T({\bf r},t)  =  \sum\limits_{\bf k}
\left( c_{\bf k}e^{-i\omega_{\bf k}t}\right) e^{i{\bf k}\cdot{\bf r}} \\
\lefteqn{U({\bf k},t) =} \nonumber \\
& &  \sum\limits_{\bf r}
\left( \sum\limits_{j=1}^{N_A}\sqrt{\omega_j}D_j^* c_j 
\delta({\bf r} - {\bf r}_j)e^{-i \omega_j t}\right) e^{-i{\bf k}\cdot{\bf r}}.
\end{eqnarray}
Both $T({\bf r},t)$ and $U({\bf k},t)$ are two dimensional Fourier transforms,
so in numerical
calculations the Fast Fourier Transform (FFT) can be used. The speed 
increase obtained
by using FFT instead of the direct summation is enormous especially in 
simulations with a large
number of atoms. In some simulations it can be said that 
only this method makes these simulations
possible.

There are several natural checks for the numerical simulations. 
First of all the norm of the wavefunction
has to remain unity for all times. The system is closed so the total energy
of the system
must be constant all the time. The field energy can be calculated using either
the formula (\ref{hf}) or
integrating the energy density over the whole cavity. 
The two methods should give the same results.

\subsubsection{A method to detect a local time dependent spectrum}

In the following simulations the spectrum is detected using 
so-called analyzer atoms
\cite{mhss}. Many atoms with a very small dipole coupling constant are 
put into specific
locations in the cavity. All the atoms have different transition 
frequencies in between
$\omega_{min}$ and $\omega_{max}$

\begin{eqnarray}
\omega_j=\omega_{min} + \Delta\omega\cdot (j-1) & , & \ \ \Delta\omega
=\frac{\omega_{max} - \omega_{min}}{N-1}, \nonumber \\
 & & j=1,2...N
\end{eqnarray}
Also the dipole constants are all different and very small

\begin{equation}
\label{dipole}
D_j=\frac{C}{\omega_j},
\end{equation}
where $C$ is a 
very small constant, typically C=0.0001 or so. The form (\ref{dipole}) 
of $D_j$ gives
the same decay constant $\Gamma$ for all the atoms because in two 
dimensions $\Gamma$ is directly
proportional to the product $D_j^2\omega_j^2$. Because the dipole 
coupling is small,
the atoms have very small decay constants and linewidths and only the 
radiation which is exactly on resonance with
the atom can excite it. Therefore the excitation of the atoms as 
function of $\omega$ can be
interpreted as a spectrum of the field at the position of the atoms. 
Because the interaction between the radiation and the atoms is small, the state
of the field does not change appreciably.
The method can be used to detect the local time dependent spectrum.
Two-time averages, usually used in spectrum calculations, are not needed.
A more detailed
description of the method and comparisons with the time dependent spectra 
defined using two-time averages \cite{eberly} can be found in the paper by
M.Havukainen and S.Stenholm \cite{mhss},
where it was used to detect the spectrum of a radiation emitted by a laser
driven three level atom.

\section{Simulations}

In this chapter the results of several simulations are presented. 
First we show that the energy
density profile of the free photon does not preserve its 
shape if $\omega_0$ is small as was
explained earlier. In the second and third simulation, atoms 
are used as mirrors and
beam splitters. Using these components it is possible to build 
many optical systems.
We present an interferometer as an example. 
We also present a simulation of  a two-slit experiment.
Finaly, we also briefly study a spontaneous decay of a two level
atom into the vacuum of electromagnetic modes in the two-dimensional
cavity.

\subsection{A free photon}

\begin{figure}
\centerline{\epsfig{file=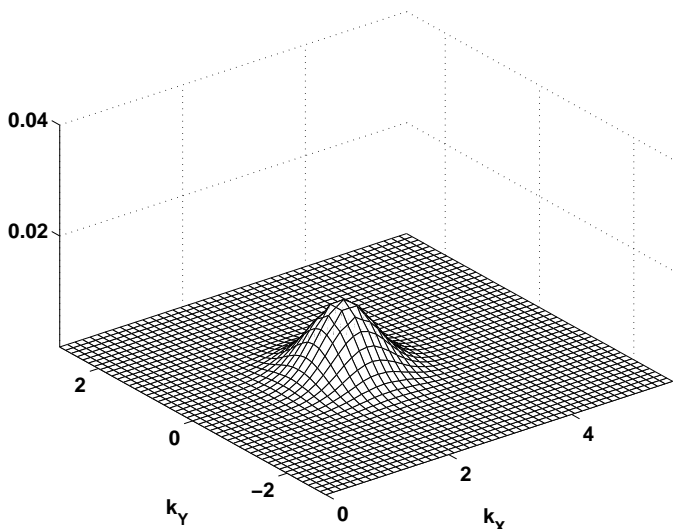,width=9cm}}
\caption{\narrowtext The probabilities $|c_{\bf k}|^2$ of the Gaussian initial state 
in ${\bf k}$-space. The parameters in the equation (\protect\ref{Gauss}) are
$x_0=-8.0$, $y_0=0.0$, $k_{x0}=4.0$, $k_{y0}=0.0$, $\Delta_{kx}^2=1.0$ and 
$\Delta_{ky}^2=1.0$. Here we consider the size of the cavity to be $L=10\pi$
and we take into account $256\times 256$ modes of the electromagnetic
field. Only one polarization ($s=1$) is taken into
account.
\label{freekspace}}
\end{figure}

\begin{figure}
\centerline{\psfig{file=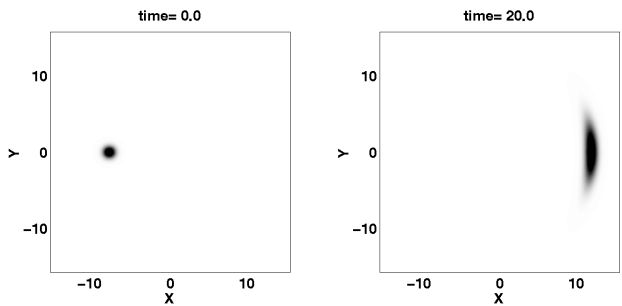,width=7.5cm,bbllx=1cm,bblly=1cm,bburx=21cm,bbury=27cm,angle=-90,clip=}}
\caption{\narrowtext
The time evolution of the energy density of 
of the initial Gaussian photon in free space.
The parameters are the same as in Fig. \protect\ref{freekspace}. 
We see that the initial wave packet (a) is nicely localized in
the configuration space while at later times 
it does not preserve its initial  shape - we see (b) the spreading of
the original wave packet in the $y$-direction.
\label{figFree}}
\end{figure}

In the first simulation the time evolution of the free photon wave packet
is studied. 
The initial
wave packet is Gaussian (\ref{Gauss}) with parameters $x_0=-8.0$, $y_0=0.0$, 
$k_{x0}=4.0$, $k_{y0}=0.0$
and $\Delta_{kx}^2=\Delta_{ky}^2=1.0$. The probabilities $|c_{\bf k}|^2$ 
of the field modes are shown 
Fig. \ref{freekspace}. The central frequency of the photon wave packet
is so small that the ${\bf k}$-vectors of the modes
with nonzero amplitudes are not parallel. We would expect this to
 be observed as was explained
earlier. The time evolution of the energy density at two time values 
is shown in
Fig. \ref{figFree}. The initial Gaussian photon wave packet has an 
energy density centered at $x=-8.0$, $y=0.0$. 
The wave packet is  moving to the right. During the free evolution 
energy density  becomes delocalized.
From the figure we see that 
at $t=20.0$  the width in the $y$-direction is much larger 
than the initial value. This spread of the width of the original wave packet
is a standard quantum-mechanical effect.

\vspace*{\fill}
\pagebreak

\subsection{A mirror}

It is possible to ``build'' mirrors and beam splitters using two 
level atoms. In the next simulation
many atoms with large dipole constants were arranged into a slab configuration.
We take a $45^o$ angle between
the slab and the $x$-axis. 
We assume all atoms to have the same transition frequencies and dipole
constants.
The initial photon wave packet 
has a Gaussian distribution (\ref{Gauss}) with
parameters $k_{x0}=5.0$, $k_{y0}=0.0$ and $\Delta_{kx}^2=\Delta_{ky}^2 = 0.125$. 
The atoms in the slab are exactly on resonance with the incoming 
photon wave packet (i.e. the central frequency of the wave packet 
$\omega_0=5.0$ is equal
to the transition frequency of the atoms). 
The dipole constant
is large $D=0.5$. 
We assume that the mirror is composed of 
eight layers of atoms as close to each other as possible. 
In our case we assume to distance between neighboring layers of the atoms 
to coincide with the grid in the configuration space (the grid spacing is
 $\Delta x$ and for the given orientation of the
mirror the distance between the different atomic layers is chosen to be
$\Delta X=\sqrt{2}\cdot\Delta x=0.17$).
The central  wavelength of the incoming photon wave packet 
 is $\lambda=1.26$ so the
difference between the neighboring atoms much shorter than 
the wavelength of the incoming wave packet.

\begin{figure}
\centerline{\psfig{file=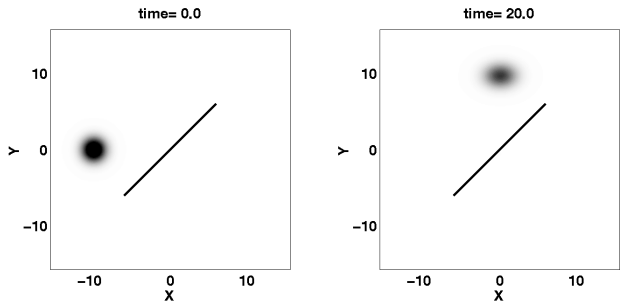,width=7.5cm,bbllx=1cm,bblly=1cm,bburx=21cm,bbury=27cm,angle=-90,clip=}}
\caption{\narrowtext
The energy density of the one-photon wave packet reflected
by a mirror composed of two-level atoms.
The initial photon is Gaussian (\protect\ref{Gauss}) with parameters
$x_0=-8.0$, $y_0=0.0$, $k_{x0}=5.0$, $k_{y0}=0.0$, 
$\Delta_{kx}^2=0.125$ and $\Delta_{ky}^2=0.125$.
The atoms of the mirror are exactly on resonance
with the central frequency
of the  photon wave packet ($\omega=5.0$). 
The dipole constant of the atom is chosen to be 
$D=0.5$. The total number of atoms considered in this simulation 
was 1584. 
The number of modes is the same as in the simulation
presented in Fig. \protect\ref{figFree}.
\label{figMirror}}
\end{figure}

We plot the energy density of the one-photon wave packet reflected by
the mirror in Fig. \ref{figMirror}. 
Firstly we plot the initial wave packet at $t=0.0$ (a).
The photon is coming towards the atoms of the mirror. These atoms
become excited by the incoming wave packet. The ``secondary'' radiation
which is emitted by the atoms interfere with the incoming wave packet.
This secondary radiation can formally be expressed as a sum of the two
terms - the first destructively interfere with the incoming wave
packet. As a consequence of this interference the incoming wave packet
is ``destroyed'' (i.e. becomes extinct).
 The other part of the radiation which is ``collectively''
radiated by the atoms of the mirror represents the reflected wave packet.
In fact, the process of reflection of the wave packet by atoms of the mirror
represents purely quantum (microscopic) version of the Ewald-Oseen 
extinction theorem \cite{Born}.
In Fig.~\ref{figMirror}(b) we have chosen conditions such that at  
 $t=20.0$ all the radiation is reflected by the atoms.

\vspace*{\fill}
\pagebreak

The direction of propagation of the reflected wave packet is the same
as expected in the classical theory. The energy density
 compared to the incoming wave packet is changed but is still
clearly localized.
Note that the energy density is not perfectly symmetrical. 
The reason is the same as in the simulation with a free photon, i.e.  
the distribution in $k$-space is broad and near the origin so the
spread of the wave packet is clearly seen. Additionally, the interference
between components of radiation emitted by different atoms of the mirror
plays a role. 
In the
left part of Fig. \ref{figPhotonatMandBS} we see the 
energy density of the photon wave packet close to the surface of the
mirror. We see that the 
incoming and reflected parts interfere.
We also see that no energy is transmitted by the atomic slab. 
In this sense the
atoms serve as a mirror. Nevertheless, one has to remember that
the atoms during the process of reflection of the original wave packet
become excited, that is the mirror under consideration has its own 
``internal'' (quantum) degrees of freedom, so the part of the original
energy can be (transiently) absorbed by the mirror. This also result
in the fact that this quantum mirror might become entangled with the reflected
wave packet.

We note that 
the parameters of the atoms in this simulation were carefully chosen
in such a way that the atoms really form a mirror. If the parameters
are changed then part of the radiation can be transmitted, that is the
collection of the atoms can play a r\^ole of a beam splitter.

\subsection{A beam splitter}

\begin{figure}
\centerline{\epsfig{file=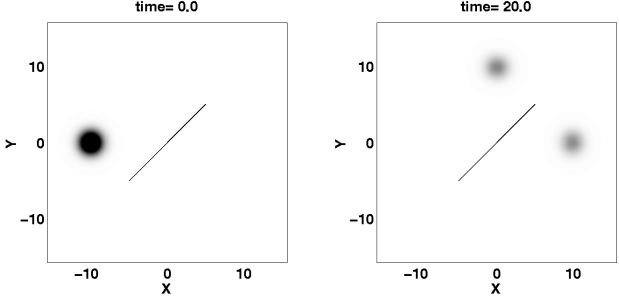,width=7.5cm,bbllx=1cm,bblly=1cm,bburx=21cm,bbury=27cm,angle=-90,clip=}}
\caption{\narrowtext
The energy density of the photon wave packet which is
splitted by a quantum beam splitter composed of a set of  two level atoms
composing a one-dimensional crystal - the quantum beam splitter.
The initial photon is Gaussian (\protect\ref{Gauss}) with parameters
$x_0=-10.0$, $y_0=0.0$, $k_{x0}=15.0$, $k_{y0}=0.0$, $\Delta_{kx}^2=0.125$ 
and $\Delta_{ky}^2=0.125$.
The transition frequency $\omega=10.4$
of the atoms is detuned from the central
frequency of the incoming photon wave packet $\omega_0=15.0$.
The total number of atoms is equal to 881, while 
the number of modes is the same as in the simulation
presented in Fig. \protect\ref{figFree}.
Here we again assume the 
 dipole constant of the atoms to be  $D=0.5$.
\label{figBS}}
\end{figure}

In the previous simulation we have  shown that it is possible to build 
an almost perfect mirror
using two level atoms, assuming the parameters are chosen correctly. 
Using slightly different parameters,
we find that the atoms can behave as a beam splitter. 
There are several ways how to modify the ``mirror'' configuration
to obtain a beam splitter -- for instance, we can consider a
smaller number of atoms, or we can 
 decrease the dipole constants, or   change the resonance 
frequencies of the atoms.
We tried all the possibilities and the most satisfactory results were 
obtained by detuning the atoms. 
The frequencies of the atoms are now taken to be $\omega=10.4$.
The center frequency of the incoming photon wave packet  is
$\omega_0=15.0$, i.e. the detuning is really large. 
The time evolution of the energy density of the electromagnetic field 
in this case is
shown in Fig. \ref{figBS}. The line in the middle represents the positions of 
the detuned atoms. There
is only one layer of atoms instead of eight as in the mirror simulation. 
At $t=0.0$ the photon is propagating towards the atoms. 
Here again the incoming wave packet excites the atoms. 
Now the quantum interference between the incoming and emitted radiation
is such that part of the original wave packet is transmitted by the
layer of the atoms. The other part is reflected.
In Fig.~\ref{figBS} we clearly see that at
 $t=20.0$ the original wave packet is 
 split into two parts
propagating up and to the right. The energy is divided equally, that is 
the atoms form a 50-50 beam splitter for the incoming
photon. Here we stress that the beam splitter under consideration
has its own internal degrees of freedom and transiently it becomes
excited. Nevertheless, after a while the atoms completely emit the
excitation energy and the beam splitter is in a ground state - at this
point it is completely disentangled from the one-photon 
radiation field which
is now in a pure superposition state with two macroscopically
distinguishable components (reflected and transmitted).

\vspace*{\fill}
\pagebreak

\vspace*{-2cm}

\vspace{-1cm}

\begin{figure}
\centerline{\psfig{file=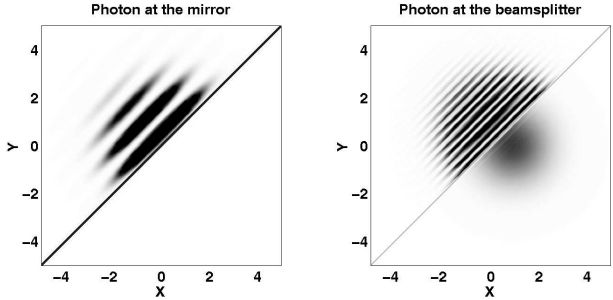,width=7.5cm,bbllx=1cm,bblly=1cm,bburx=21cm,bbury=27cm,angle=-90,clip=}}
\caption{\narrowtext
The energy density of the electromagnetic field at the moment when
the incoming wave packet interfere with the radiation re-emitted
by the atoms of the mirror (left) and the beam splitter (right).
The central  wavelength of the photon wave packet in the case of the 
mirror simulations is taken to be longer compared to the case of the
beam splitter simulations. The interference pattern in the two cases
is different. We see that in the case of the beam splitter part of
the radiation is transmitted.
The parameters of the simulations 
are specified in previous figures.
\label{figPhotonatMandBS}}
\end{figure}

In the right hand part of Fig.~\ref{figPhotonatMandBS} 
energy density of 
the photon wave packet is shown close to the ``surface'' of the 
beam splitter. 
To the left of the atoms the incoming
and reflected wave packets 
 interfere. We also see that a fraction of the original
radiation is able to ``pass'' the atoms and to 
continue to propagate to the right. 
The wavelength of the photon was chosen shorter than in the
mirror simulation, which can be seen from the interference structure.

We have also studied spectral properties of reflected and transmitted
parts of the original wave packet.
In this situation 200 atoms were used to
detect the time dependent spectra of the two outgoing parts of the photon by
applying the method described earlier. Both
spectra were identical to the spectrum of the incoming photon. 
This means that  our quantum 
beam splitters and mirrors are linear devices, which is important if
we want to build optical networks out of the considered optical
elements.

\subsection{Parabolic mirror}
\begin{figure}
\centerline{\psfig{file=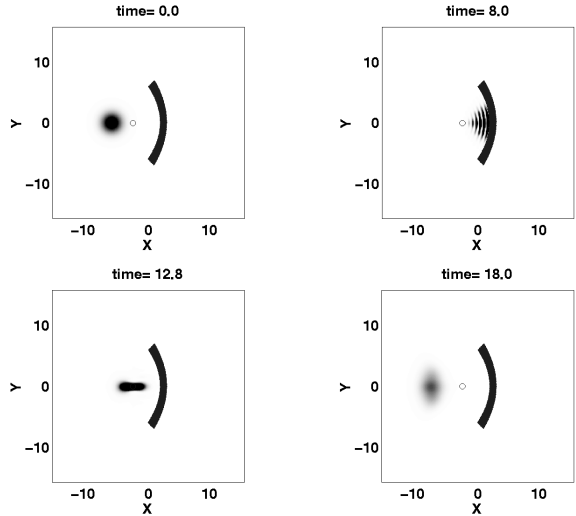,width=7.5cm,bbllx=1cm,bblly=1cm,bburx=21cm,bbury=27cm,angle=-90,clip=}}
\caption{\narrowtext
The time evolution of the energy density of the one-photon wave packet
reflected by a parabolic mirror composed of two-level atoms.
At time $t=0$ the wave packet is localized to the left of the focus of the
mirror (a little circle in the figure) and propagates towards it.
At time $t=8.0$ we see an interference pattern due to interference between
the incoming wave packet and the re-emitted radiation. 
At time $t=12.8$ the original wave packet is completely 
reflected by the mirror and is localized around the focus.
 The spatial dependence of the energy density
is determined by the shape of the mirror. We can observe a 
reduction of the width of the reflected wave packet in the $y$ direction.
At time $t=18.0$ the wave packet is spread significantly. We see that
the maximal energy density is now smaller than in the original wave packet
(compare with figure $t=0.0$).
The number of atoms from which the parabolic mirror is composed is
$N_A=1100$. The parameters of the atoms are the same as in
Fig. \protect\ref{figMirror}. The initial photon is Gaussian
(\protect\ref{Gauss}) with parameters $x_0=-6.0$, $y_0=0.0$,
$k_{x0}=5.0$, $k_{y0}=0.0$, $\Delta_{kx}^2=0.125$ and $\Delta_{ky}^2=0.125$.
\label{figParabola}}
\end{figure}

Another  illustration of the power of our microscopic model of
optical elements is the parabolic mirror. In fact, 
it is possible to ``build'' out of two-level atoms 
 mirrors of arbitrary shapes. 
In the next simulation, the photon wave packet is
propagating towards a parabolic mirror the shape of which is described
by the equation
$x=x_0 + \frac{1}{2p}y^2 = 2 - \frac{1}{18}y^2$.
The focus of the parabola 
is at the point $x=x_0+\frac{p}{2} = -2.5$, $y=0$. 
The time evolution of the energy
density is shown in Fig. \ref{figParabola}. At $t=0.0$ 
the Gaussian photon is propagating
towards the parabola. The little circle in between the photon 
and the parabola shows the
position of the focus. At $t=8.0$ we see the photon wave packet
being  reflected from the parabola. We see the interference between
the incoming and the re-emitted radiation. 
We note that at $t=12.8$
 most of the radiation goes through the focus. In the last
figure ($t=18.0$) the photon wave packet propagates to the left. 

\vspace*{\fill}
\pagebreak

\subsection{Interferometer}

\begin{figure}
\centerline{\psfig{file=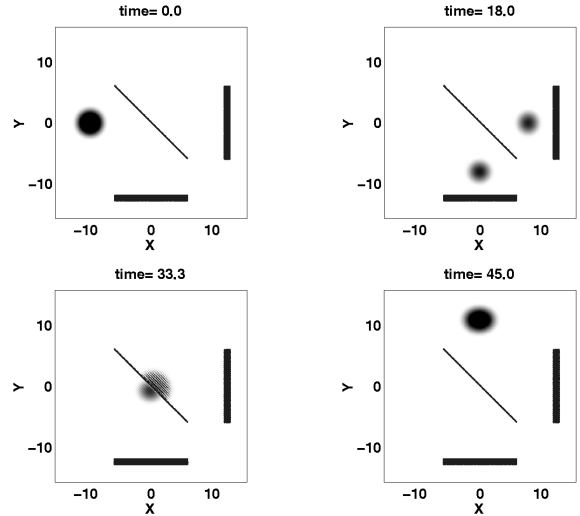,width=7.5cm,bbllx=1cm,bblly=1cm,bburx=21cm,bbury=27cm,angle=-90,clip=}}
\caption{\narrowtext
The time evolution of the energy density of the
one-photon wave packet in an interferometer. 
The distance of the two mirrors from the
beam splitter is the same. 
We see that the interference results in a wave packet propagating upwards.
The initial wave packet, the beam splitter and the mirrors have the
same parameters as in figures considered above.
Here the mirrors and the beam splitter are specified in 
Figs.~\ref{figMirror} and \ref{figBS}, respectively. 
\label{figIFshift00}}
\end{figure}

Using a quantum beam splitter and two quantum  mirrors we can 
``construct''
a single-photon interferometer (see  Fig.~\ref{figIFshift00}). 
Here the one-photon wave packet  comes
towards the beam splitter ($t=0.0$)
and is divided into two parts which 
propagate towards the mirrors ($t=18.0$). 
The distances
 of the mirrors from the beam splitter are exactly the same.
The mirrors reflect the radiation back to the beam splitter. 
At $t=33.3$ the two reflected parts  reach the beam splitter. 
Each wave packet considered individually would be splitted by the
beam splitter into two parts going left and up (i.e. transmitted
and reflected).

\begin{figure}
\centerline{\psfig{file=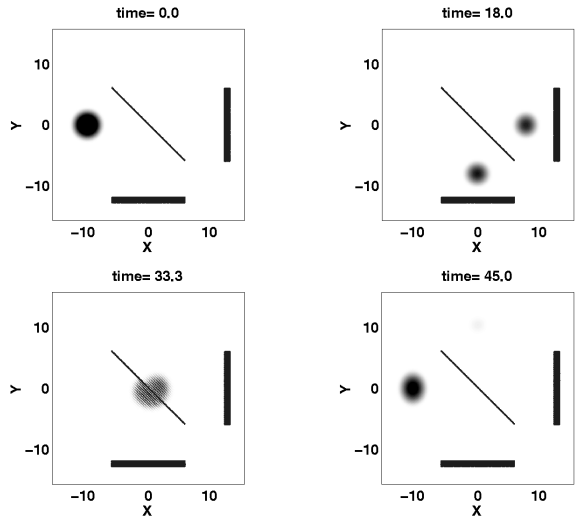,width=7.5cm,bbllx=1cm,bblly=1cm,bburx=21cm,bbury=27cm,angle=-90,clip=}}
\caption{\narrowtext
Same as in Fig. \protect\ref{figIFshift00} except 
the distances of the
mirrors from the beam splitter differ by one wavelength. 
This difference leads to a quantum interference which results
in a wave packet propagating to the left.
The initial wave packet, the beam splitter and the mirrors have the
same parameters as in figures considered above.
\label{figIFshift043}}
\end{figure}

On the other hand 
due to the quantum interference between the components of the radiation
field coming from the two mirrors
we can observe something completely different:
If the optical paths of the two components are equal then their 
relative relative phase  is such that quantum interference
results in an emergence of a single-photon wave packet 
traveling up ($t=45.0$). 
On the contrary, if the distances of the two mirrors from the beam splitter
are not equal then the relative phase of the two components which interfere
on the beam splitter after being reflected by the mirrors can result
in a wave packet traveling left (see  Fig.~\ref{figIFshift043}). 
Here the difference of the optical paths is
approximately 
 one central wavelength of the original wave packet.
We see that in this case most of the energy travels in a form of  a 
wave packet to the left.

\vspace*{\fill}
\pagebreak

\vspace*{-0.5cm}

\begin{figure}
\centerline{\psfig{file=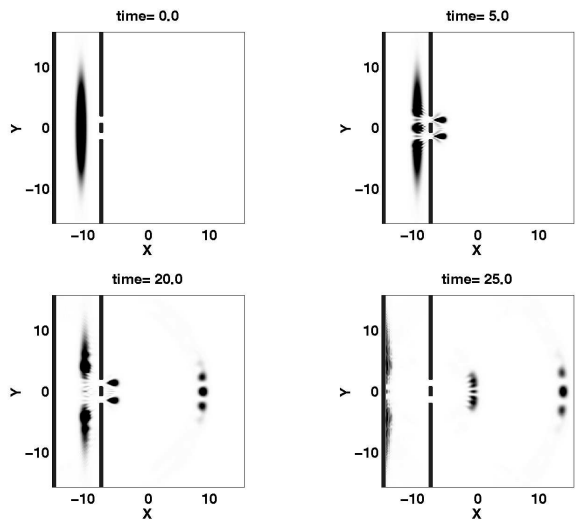,width=7.5cm,bbllx=1cm,bblly=1cm,bburx=21cm,bbury=27cm,angle=-90,clip=}}
\caption{\narrowtext
The time evolution of the energy density of a photon wave packet
in the two-slit experiment. A ``plane'' wave packet propagates
towards the mirror with two slits. The mirror is composed of two level atoms
as in previous figures except in this case there are two slits now.
Another quantum mirror is considered to be located to the left of the
two-slit mirror. This configuration is chosen to make sure that none of the
original energy passes to the right due to the periodic conditions 
we imposed on the Schr\"{o}dinger equation.
To make the figure more transparent we use a logarithmic 
 scale for the energy density of the field. The number of modes in this
simulation is $512\times 512$. The total number of atoms used in
the mirrors is $N_A=7872$.
\label{figTwoSlit}}
\end{figure}

\subsection{Two-slit experiment}

The microscopic quantum model we study in this paper can be 
 also used to study the two-slit experiment.
Let us assume  the photon wave packet
which has a very broad energy density in the $y$-direction, i.e. this
wave packet models a plane wave which
approaches the mirror with two slits, see Fig.~\ref{figTwoSlit}.
On the left we have placed another mirror. Without
it, the part of the plane wave which is reflected from the double slit 
mirror would
disappear at the left and reappear on the right because 
of the periodic boundary conditions we use in our simulations.

The original one-photon  
wave packet ($t=0.0$) propagates towards the mirror with two slits.
At $t=5.0$ the ``plane'' wave packet  is reflected from  mirror. 
Some of the energy propagates through the slits (i.e. there is
a nonzero probability that the original one-photon wave packet can
be transmitted through the mirror via the slits).
We see that through each slit a part of the energy propagates to the
right - the interference between these components of the electromagnetic
field are clearly seen (see $t=20.0$ and $t=25.0$).

The ``plane'' wave packet which   has been reflected by the mirror with
two slits is then reflected by the left mirror and then again by the 
the double slit mirror. Here  
part of the energy ``goes'' through
the slits again forming a second, more complex, interference pattern
(see $t=20.0$). This process of bouncing of the original wave packet between
two mirror continues and each time a fraction of the energy passes through
the slits.

\begin{figure}
\centerline{\psfig{file=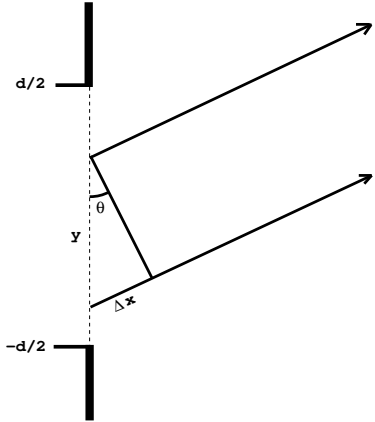,width=7.5cm,bbllx=1cm,bblly=1cm,bburx=21cm,bbury=27cm,clip=}}
\caption{\narrowtext
We show one slit in the mirror. 
The width of the slit is $d$. The optical path difference
of the radiation coming from two different points at the slit is $\Delta x$.
\label{slitfigure}}
\end{figure}

It is interesting to compare the interference structure with the theoretical
prediction derived within classical optics. 
The formula can be calculated using Huygens'
principle \cite{moller}. According to this, every point at the slit can be
considered to be a source of secondary wavelets. The total intensity profile
is a superposition of these wavelets. Let us first consider a text book 
treatment of a one slit mirror,
Fig. \ref{slitfigure}. 

\vspace*{\fill}
\pagebreak

The plane wave packet of the 
frequency $\omega$ is coming from
the left towards a slit of a width $d$. The field strength on the right coming
from a specific point at the slit is
proportional to the phase factor $e^{i(kx-\omega t)}$. The phase difference
of the radiation coming from two different spatial points in the slit is
$\Delta x=y\sin\theta$, if the distance from the mirror is long enough.
According to the Huygens' principle the total radiation is a superposition

\begin{eqnarray}
E & \propto & e^{i(kx-\omega t)}\int\limits_{-d/2}^{d/2}e^{iky\sin\theta}dy 
\nonumber \\
  & = & e^{i(kx-\omega t)}\frac{2}{k\sin\theta}\sin(\frac{kd}{2}\sin\theta).
\end{eqnarray}
For two slits of width $d$ and a separation $a$ we have two integrals

\begin{eqnarray}
E & \propto & e^{i(kx-\omega t)}\left( \int\limits_{-d/2}^{d/2}e^{iky\sin\theta}dy
+\int\limits_{-a-d/2}^{-a+d/2}e^{iky\sin\theta}dy \right) \nonumber \\
  & = & e^{i(kx-\omega t + \frac{ka}{2})}
\frac{\cos(\frac{ka}{2}\sin\theta)\sin(\frac{kd}{2}\sin\theta)}{k\sin\theta},
\end{eqnarray}
which gives for the intensity

\begin{equation}
\label{theoreticalintensity}
I\propto E^*E=\frac{\cos^2(\frac{ka}{2}\sin\theta)\sin^2(\frac{kd}{2}\sin\theta)}{(k\sin\theta)^2}.
\end{equation}

\begin{figure}
\centerline{\epsfig{file=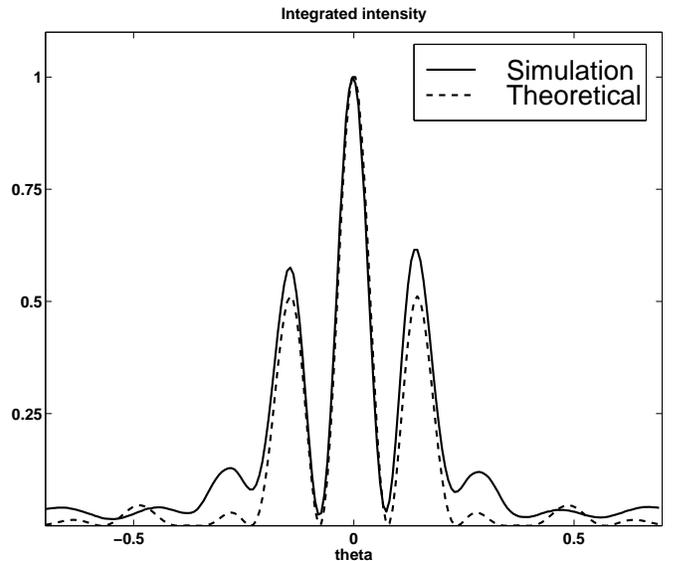,width=7.5cm,angle=90}}
\caption{\narrowtext
We present the intensity of the radiation field
at the region behind the two-slit mirror. The dashed line
corresponds to the intensity derived from a classical
model [see Eq.(\ref{intensitytheta})] while the solid line
is obtained from our numerical simulations based on purely
quantum description of the process. 
We see a very good agreement between the two
results for small $\theta$. 
\label{figThetaIntensity2curves}}
\end{figure}

On the other hand, we can use results of our numerical simulations
and evaluate 
the intensity of the radiation which has been created to the right of the
two-slit mirror during the first reflection of the original wave packet:

\begin{equation}
\label{intensitytheta}
I(\phi)=\int\limits_{r_{min}}^{L/2}I(r,\phi)dr.
\end{equation}
To neglect the contribution of the second reflection we 
take the lower bound of the integral over the polar coordinate
$r$ to be $r_{min}=10$. 
The theoretical prediction (\ref{theoreticalintensity}) and 
the intensity derived from our 
 simulations (\ref{intensitytheta}) are shown in Fig.
\ref{figThetaIntensity2curves}.
Both intensities are normalized in such a way that their maximum 
is equal to unity. Near $\theta=0$
the agreement between  the two results is very 
good. For larger values of $\theta$ there is a difference between
the two lines which is understandable because we are comparing
a classical result with a numerical simulation of a quantum model
with realistic features such as the nonzero width of a mirror composed
of two-level atoms or the wave packet which is not a plane wave, etc.
Taking these differences into account 
it is surprising that the two pictures coincide so well.

\subsection{Decay of a two level atom}
\begin{figure}
\centerline{\psfig{file=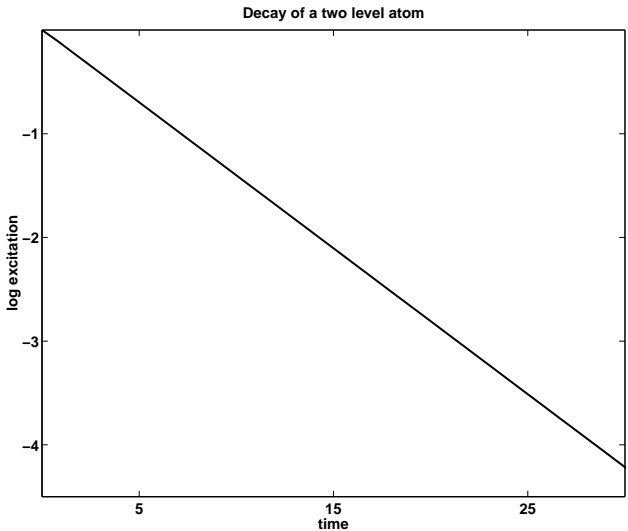,width=7.5cm,angle=90,bbllx=1cm,bblly=1cm,bburx=21cm,bbury=27cm,clip=}}
\caption{\narrowtext
The exponential decay of an excited atom. 
The atomic transition frequency and dipole
constants are $\omega=15.0$ and $D=0.05$, respectively. 
The atom is positioned in the center of the square cavity of
the linear dimension $L=10\pi$.
We consider $256\times 256=65536$ modes of the electromagnetic field.
The probability to find the atom in the excited state is plotted
in the logarithmic scale - the exponential character of the decay
is clearly seen. The corresponding decay rate is $\Gamma=0.14$.
\label{decayexcitation}}
\end{figure}

\begin{figure}
\centerline{\psfig{file=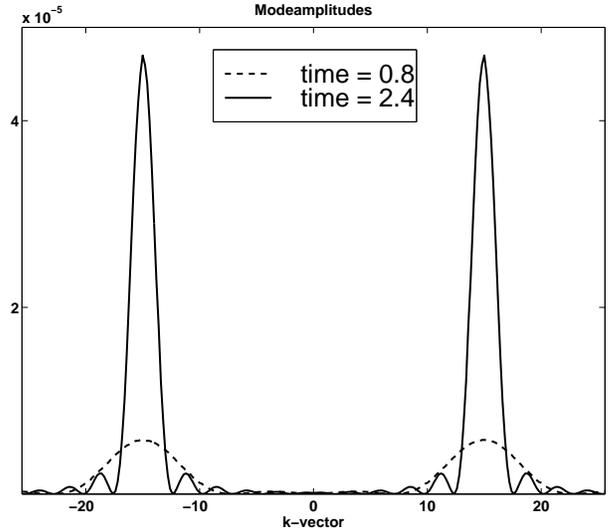,width=7.5cm,bbllx=1cm,bblly=1cm,bburx=21cm,bbury=27cm,angle=90,clip=}}
\caption{\narrowtext
We present probabilities of the excitation of the modes
with $k_y=0$. We see that for times large enough the
modes with $|k_x|=15$ are dominantly excited, i.e. the field
mode at frequencies close to the resonant transition frequency
of the atom are most excited.
\label{decaykslides}}
\end{figure}

Till now we have considered in our simulations that the field has been
initially excited and all the atoms were initially in the ground state.
Obviously, our model can be also applied to a situation when one of the
atoms is excited and the field is initially in the vacuum state 
(i.e., we still restrict ourselves to the one-excitation subspace of
the total Hilbert space).
In this section we briefly discuss the problem of a spontaneous decay
of a two-level atom in a two-dimensional cavity. We consider the atomic
transition frequency to be 
$\omega=15.0$ and the dipole constant is $D=0.05$.
The atom is situated at the origin ($x=0.0$, $y=0.0$) of the 
two-dimensional cavity. The number of the field modes is 
256$\times$256=65536.
In Fig. \ref{decayexcitation} we present 
the natural logarithm of the excitation probability of the atom as a 
function of time. From here we can conclude that 
the decay of the atom is approximately exponential
with a decay constant $\Gamma \simeq 0.14$. 
In Fig.~\ref{decaykslides} we present 
probabilities of the excitation of   the modes  $k_x$ ($k_y$=0).
Because the direction of the constant dipole vector of the atom is  chosen
to be in the $z$-direction, 
the amplitude profile is the same on any line which goes
through the origin of the momentum space.
As expected for times large enough 
the modes with $|k_x|=15$ are dominantly excited. 
In fact the peaks are  not exactly at the resonance frequency, there 
is a small shift which is identified to be a Lamb shift (from the figures
we cannot see this but the shift can be determined from  numerical values
obtained in the simulation).

\begin{figure}
\centerline{\psfig{file=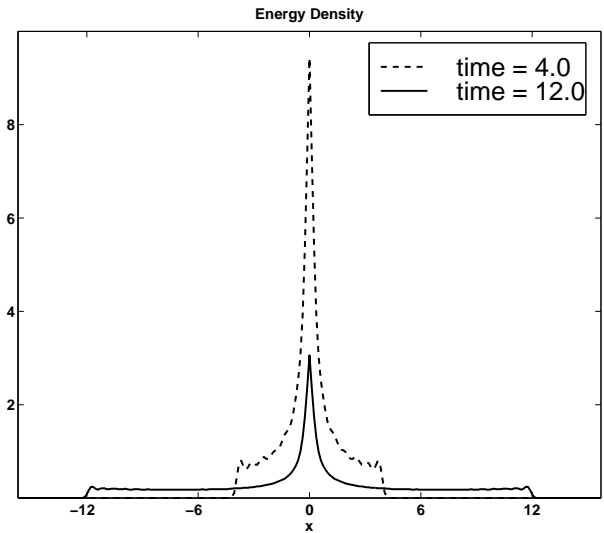,width=7.5cm,bbllx=1cm,bblly=1cm,bburx=21cm,bbury=27cm,angle=90,clip=}}
\caption{\narrowtext
We plot the 
energy density of the one-photon
wave packet emitted by the decaying atom. 
The parameters of the atom are the same as in Fig. \protect\ref{decayexcitation}}
\label{decayintensity}
\end{figure}

We plot the energy density of the one-photon wave packet emitted by
the decaying atom in  Fig. \ref{decayintensity}. Because of the rotational 
symmetry of the problem we plot just one ``cut'' ($y=0$) in the energy density
as a function of $x$.
The energy density is presented for 
two times $t=4.0$ and $t=12.0$. At both times there
is a peak in the center where the atom is positioned. This means
that at these two times the atom still emits the radiation  
(which is in agreement with the chosen decay rate $\Gamma=0.14$).
We turn our attention to the fact that 
at $t=4.0$ the energy density 
is nonzero only for 
$|x|\leq 4.0$. Analogously for the time 
$t=12.0$ the energy density is nonzero only for  $|x|\leq 12.0$. 
This reflects the fact that the causality is preserved in our simple
quantum-mechanical treatment of the decay of the two-level atom
in the cavity. 
Here we have presented just few features of the decay, the complete
description of the process deserves more detailed discussion.
For instance, one might be interested on how the decay depends on
the mode spectra, the position of the atom, what are the values of
the Lamb shift, how the decay depends on the frequency 
cut-off, etc. We will address these questions elsewhere.

\section{Conclusion}

We have shown the results of many quantum mechanical simulations with two 
level
atoms and a one photon wave packet 
 inside a two dimensional cavity. The initial
basis
vectors are restricted to admit only one excitation. Because a rotating wave
interaction between the radiation and the atoms is used, basis vectors
with more excitations acquire no excitation. For these kinds of states
the special numerical technique which utilizes FFT (Fast Fourier Transform)
may be used. Using FFT the simulations become orders of magnitude faster,
allowing more modes and atoms to be included.

The atoms are at fixed positions and it is possible to build complicated
structures with different kinds of atoms.
Several layers of atoms which are on resonance with the incoming
radiation form a quantum mechanical mirror if the density of the atoms is
high enough. The mirror may have an arbitrary shape. In our simulations
usual flat and parabolic mirror were used.
One layer of detuned atoms forms a beam splitter. We have shown
that, using mirrors and beam splitters, it is possible to build complicated
optical networks. As an example the time evolution of a photon in an
interferometer was studied.

Usually the optical components are taken to be classical objects which
give boundary conditions to the quantum mechanical time evolution or
determine the modes used in a quantization. In our simulations the whole
system including beam splitters and mirrors is in a well-defined quantum
mechanical state. In addition to the simulations shown in this paper is is
possible to build more complicated networks of beam splitters and mirrors.
One interesting possibility is to build cavities of arbitrary shape and
study the time evolution of the photon intensity inside the cavity. It
is also possible to use moving atoms in the simulations allowing moving
beam splitters and mirrors to be built. One extension of the current model
would be to take basis states with more than one photon excitation into 
account.
However, the number of basis states with a given excitation increases so 
rapidly
that it is unlikely to be possible to use the methods of this paper for
fields of higher intensity. Thus all the phenomena of nonlinear optics
require novel computational approaches.

\section{Acknowledgements}

We want to thank the Academy of Finland and the Slovak Academy of Sciences
for the financial support. This work was supported in part by the 
Royal Society.
Computers
of the Center for Scientific Computing (CSC) were used in the simulations.
In many computer programs the C++ class library ``blitz'' developed by
Todd Veldhuizen was used (http://monet.uwaterloo.ca/blitz/ ).
Finally we want to thank K-A. Suominen for reading the preprint and
correcting several misprints.

\end{multicols}

\newpage


\begin{thebibliography}{99}

\bibitem{Loudon}
 {\sc Loudon, R.}, 1983, {\em The Quantum Theory of Light}, 2nd ed.
        (Clarendon Press, Oxford).   

\bibitem{MilburnandWalls} 
   {\sc Milburn, G.J.}, and {\sc Walls, D.F}, 1994, 
{\em Quantum Optics}, (Springer, Berlin).

\bibitem{Mandelfree}  
    {\sc Mandel, L.}, and {\sc Wolf, E.}, 1995, 
    {\em Optical Coherence and Quantum Optics}, (Cambridge University Press, 
     Cambridge). 

\bibitem{Gardiner} 
   {\sc Gardiner. C.W.}, 1991,
{\em Quantum Noise}, (Springer, Berlin).

\bibitem{Steane}
  {\sc Steane, A.}, 1998, {\em Rep. Prog. Phys.} {\bf 61}, 117.


\bibitem{berman} {\em Advances in Atomic, Molecular and Optical Physics},
  Supp. {\bf 2}, {\it ``Cavity QED''}, 1994, 
  edited by P.R. Berman (Academic Press, New York).

\bibitem{weissman} 
 {\sc Weissman, Y.},  {\em Optical Network Theory}, 1992, 
 (Artech House, Boston).

\bibitem{ekert} 
  {\sc Ekert, A.K.}, and {\sc Knight, P.L.}, 1990,
     Phys.Rev. A {\bf 42}, 487; 1991, {\it ibid}, {\bf 43}, 3934.

\bibitem{stenholm94} 
    {\sc Stenholm, S.}, 1994, {\em J. mod. Optics}  {\bf 41}, 2483.

\bibitem{torma} 
    {\sc T\"orm\"a, P.}, and {\sc  Stenholm, S.}, 1995, 
 {\em J. mod. Optics} {\bf 42}, 1109.

\bibitem{stenholm95} 
    {\sc Stenholm, S.}, 1995, {\em  Appl. Phys. B} {\bf 60}, 243.

\bibitem{buzek2} 
   {\sc Bu\v{z}ek, V.,  Drobn\'{y}, G., Kim, M.G.,  Havukainen, M}
    and {\sc Knight, P.L.}, 1998, {\em Numerical simulations of fundamental
processes in cavity QED: Atomic decay}, LANL e-print archive  
{\em quant-ph/9812001}.

\bibitem{buzek1} 
   {\sc Bu\v{z}ek, V.}, 1989, {\em Czech. J. Phys. B} {\bf 39}, 345;
see also 
   {\sc Bu\v{z}ek},  and {\sc Kim, M.G.}, 1997, {\em J. Korean Phys. Soc.} 
   {\bf 30}, 413.


\bibitem{nrinc} 
  {\sc Press, W.H.,  Teukolsky, S.A.,  Vetterling, W.T.}, and 
{\sc  Flannery, B.P.}, 1995, 
 {\em Numerical Recipes in C}, 2nd ed., (Cambridge University Press, 
Cambridge), chapter {\bf 16.1}.

\bibitem{mhss} 
   {\sc Havukainen, M.}, and  {\sc Stenholm, S.}, 1998,  
   {\em J. mod. Optics} {\bf 45}, 1699.

\bibitem{eberly} 
   {\sc Eberly, J.H.}, and {\sc W\'odkiewicz, K.}, 1977, 
{\em J. Opt. Soc. Am.} {\bf 67},  1252.

\bibitem{Born}
   {\sc Born, M.}, and {\sc Wolf, E.}, 1980, {\em Principles of Optics},
   6th edn. (Pergamon Press, Oxford). 

\bibitem{moller} 
    {\sc M\"oller, K.D.}, 1988,  
{\em Optics}, (University Science Books, Mill Valley, California).



\end{thebibliography}
\end{document}